\documentclass[letterpaper]{jpconf}
\usepackage{graphicx}

\def \met {E_{T}\hspace*{-13pt}\slash \hspace*{9pt}}
\def \NI {{{\tilde{\chi}}_{1}}^{0}}
\def \NII {{{\tilde{\chi}}_{2}}^{0}}
\def \CI {{{\tilde{\chi}}_{1}}^{\pm}}
\def \C {{{\tilde{\chi}}}^{\pm}}
\def \N {{{\tilde{\chi}}}^{0}}

\begin{document}
\title{New Physics at CDF}

\author{Melisa Rossi	}

\address{Istituto Nazionale di Fisica Nucleare - Sezione di Trieste \\
(on behalf of the CDF Collaboration)}

\begin{abstract}
We present the current status of searches for physics beyond the Standard Model at the Tevatron 1.96-TeV proton-antiproton collider using data collected with the CDF experiment. We cover searches for supersymmetry, extra dimensions and new gauge bosons.
\end{abstract}

\section{Introduction}
The Standard Model (SM) of particle interactions is one of the major achievements of fundamental science. However, despite its success in predicting experimental results through the years, several fundamental issues are left open such as: the origin of dark matter in the universe, the existence of a unifying theory of all known forces and interactions, why particles have the masses we observe and the hierarchy problem.

Supersymmetry (SUSY) is one of the possible extensions of the SM. It proposes an additional symmetry that predicts the existence of an additional boson (fermion) for each SM fermion (boson). The appeal of SUSY is closely related to the fact that it naturally solves several of the open questions of the SM. Thus it is extensively tested in an experiment as CDF where many of its signatures can be searched for. On the other hand there is still no compelling reason to exclude other scenarios like Extra-dimensions and New Gauge Interaction Models, a priori.

The CDF program in physics beyond the SM is quite rich and here we present some representative searches subdivided in two main categories: SUSY and non-SUSY searches. 

\section{Supersymmetry searches}
Each SM particle is expected to be degenerate in mass with its corresponding SUSY partner. Nevertheless experimental results show that, at the energies explored by colliders so far, there is no evidence of supersymmetry yet. This leads to conclude that SUSY must be broken, even though the breaking mechanism is not known. Theoreticians have proposed several possible ways of inducing this symmetry breaking and, depending on the model chosen, one obtains different supersymmetric mass spectra which condition the strategy chosen to experimentally search for the new expected particles.
The Minimal Supersymmetric Standard Model (MSSM) is the minimal extension of the SM that includes supersymmetry in which the breaking is explicitly introduced by ad hoc operators in the Lagrangian.
Other proposed ways to break supersymmetry are the minimal supergravity scenario (mSUGRA) and the gauge-mediated SUSY-breaking scenario (GMSB).

In the most general supersymmetric theory, the derived Lagrangian includes terms which violate lepton or baryon number. These terms can be suppressed by imposing a discrete symmetry called R-parity ($R_p =  (-1)^{2s+3B+L}$). 
In a R-parity conserved scenario, SUSY particles must be pair produced. It follows that the lightest SUSY particle is stable and provides a viable candidate for cosmological dark matter. Depending on the SUSY breaking model considered, both R-parity conserved and not conserved scenarios are experimentally explored. The searches presented below focus on the first hypothesis. 

\subsection{Chargino and Neutralino associated production}

\begin{figure}[ht!]
\begin{minipage}[b]{0.55\linewidth}
\centering
	\includegraphics[width=\linewidth]{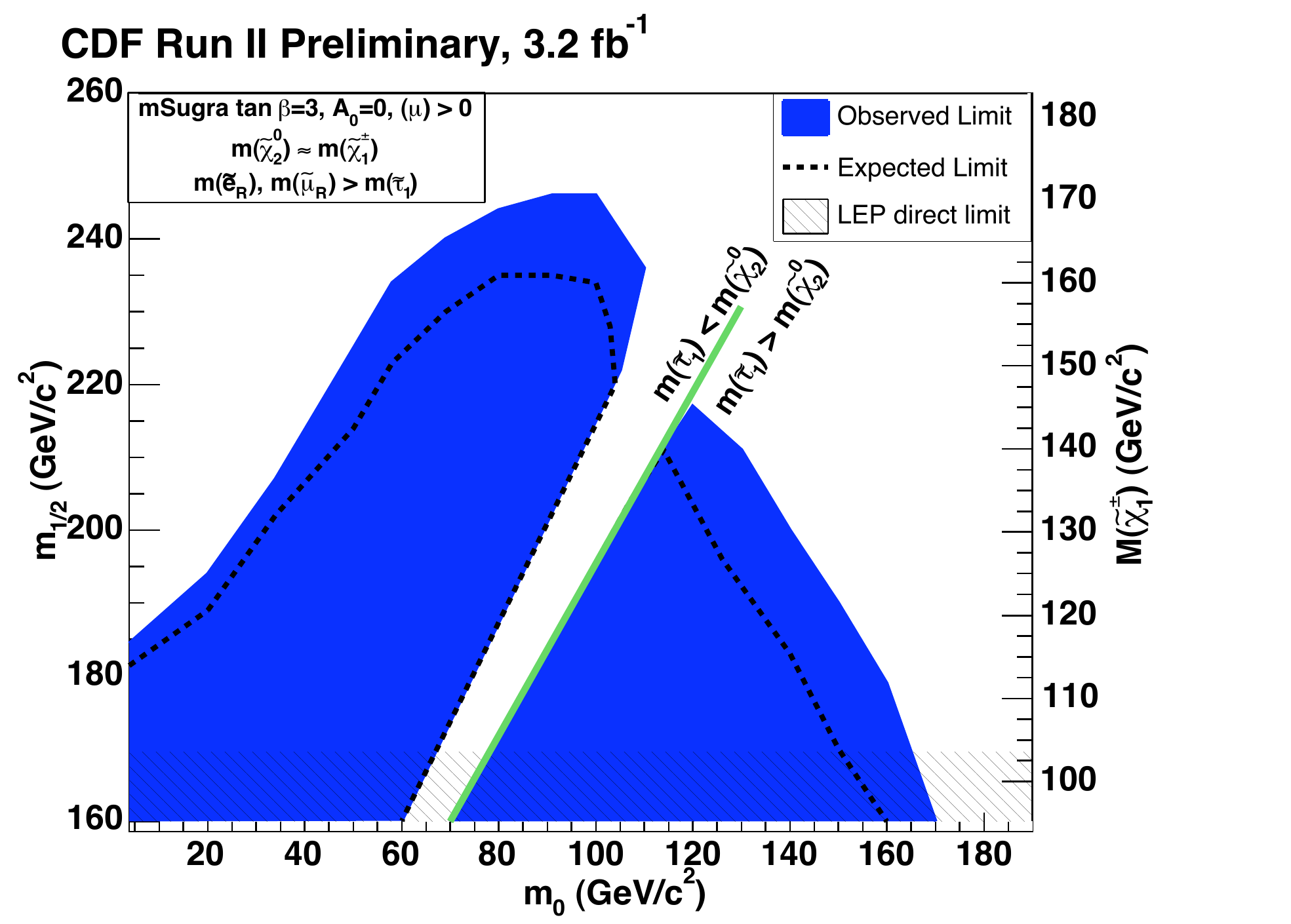}
	\caption{mSUGRA exclusion region for the $\CI\NII$ CDF search.}
	\label{fig:c1n2}
\end{minipage}
\hspace{0.02\linewidth}
\begin{minipage}[b]{0.43\linewidth}
\centering
   \includegraphics[width=\linewidth]{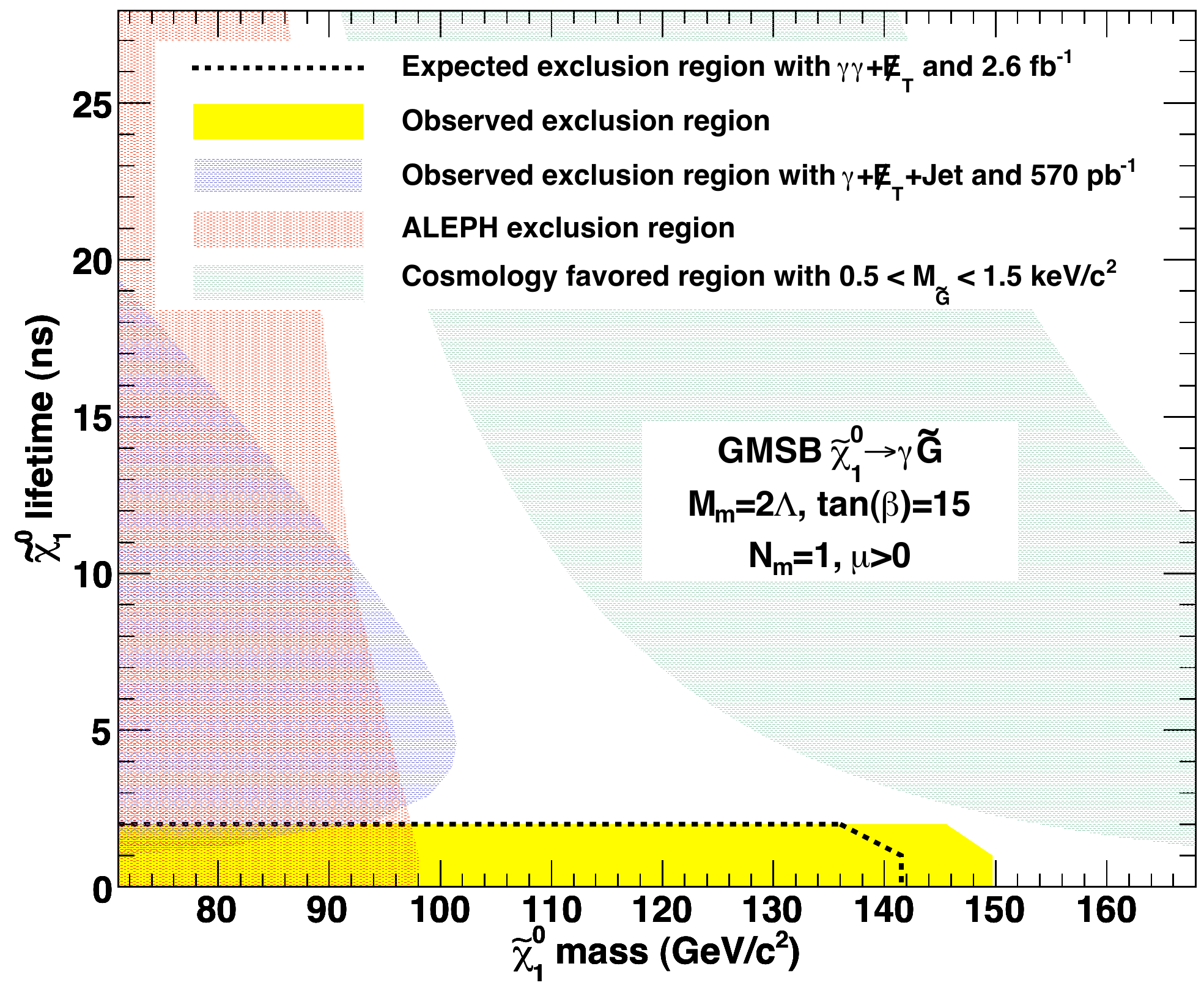} 
   \caption{Excluded regions in GMSB models for the $\gamma\gamma$ + $\met$ search.}
   \label{fig:gmsb}
\end{minipage}
\end{figure}

CDF performed a search with 3.2 $\rm{fb}^{-1}$ of data for the associated production of chargino-neutralino \cite{ref:cdfC1N2}. This is a very promising mode for SUSY discovery at hadron colliders because it leads to a final state containing 3 isolated leptons.
At reconstruction level, exclusive lepton categories are defined (electrons, muons and isolated tracks so as to recover acceptance for tau leptons) in an analysis framework which facilitates statistical combination of selected events. Control regions are defined to check agreement between expected background and the corresponding data. The control region parameter space is defined as missing transverse energy versus the highest di-lepton invariant mass in the event. Once the agreement is satisfactory, in addition to the requirement of the third lepton, some final cuts on simple kinematics variables are applied to enhance the SUSY signal fraction. Data are found to be in agreement with the expected background and results are interpreted in the context of the mSUGRA scenario (which depends on only 5 free parameters $M_0$, $M_{1/2}$, $\tan\beta$, $A_0$, $\mu$). Limits are set by fixing $\tan\beta$, $A_0$, $\mu$ and scanning the $M_0$ and $M_{1/2}$ plane as shown in Fig.\ref{fig:c1n2}. 

\subsection{$\gamma\gamma$ + $\met$ signature in GMSB models}
In the context of Gauge-Mediated Supersymmetry-Breaking Models, in which the gravitino ($\tilde{G}$) is the Lighest  Supersymmetric Particle (LSP) and the Next-To LSP is the lighest neutralino (being $\NI \rightarrow \gamma \tilde{G}$ the dominant decay), the associated production of gauginos could lead to a final state characterized by two photons and a certain amount of missing transverse energy.
The CDF Collaboration performed an analysis focusing on the $\gamma\gamma$ + $\met$ final state for low lifetime models of the $\NI$ \cite{ref:gauginos}. In addition to a 13 times larger data sample (2.6 $\rm{fb}^{-1}$), more refined methods of rejecting non-collision events and QCD background with respect to the previous CDF search for the same signature, are employed.
The approach chosen is to keep topological cuts to a minimum, while still providing useful benchmarks to compare with other experiments. After the final selection, 0 events are found which is consistent with the background estimate of 1.2$\pm$0.4 events. Exclusion regions are showed (see Fig.\ref{fig:gmsb}) and limits on GMSB models with a $\NI$ mass reach of 149 GeV$/c^2$ are set at a $\NI$ lifetime of 0 ns.

\subsection{Squarks production}

\subsubsection{Search for scalar top.}
The supersymmetric partner of the SM top quark is expected to be the lightest squark. Due to the large SM top mass, it is plausible to assume $M_{stop}<M_{top}$ as done in all CDF searches. Several stop decays are investigated at CDF such as $\tilde{t}\rightarrow \rm{b} \C$, where the chargino decays to $l\nu\N$ (being $\N$ the LSP) \cite{ref:stop2} or $l \tilde{\nu}$ (being $\tilde{\nu}$ the LSP) \cite{ref:stop3}. Here we focus on a third search: the associated production of two stops each one proceeding via a two body decay to a charm and a neutralino ($BR(\tilde{t}\rightarrow \rm{c} \NI)$ assumed to be 100$\%$ in the context of the MSSM) \cite{ref:st}.
Thus events with 2 jets plus a certain amount of missing transverse energy are analyzed where at least one of the jets is required to be tagged as originating from a heavy-flavor quark. This analysis is optimized via one Neural Network to reduce the heavy flavor multijets background plus a flavor separator to increase the fraction of c-jets. Results are obtained with 2.6 $\rm{fb}^{-1}$ of data and the achieved sensitivity allows to exclude $\tilde{t}$ masses up to 180 GeV$/c^2$ at 95$\%$ C.L. 

\subsubsection{Search for scalar bottom.}
Still in the context of the MSSM, CDF performed another analysis searching for the associated production of two SUSY partners of the bottom quark (assuming $BR(\tilde{b}\rightarrow \rm{b} \NI ) = 100\%$) \cite{ref:sb}. Thus the signature of interest is constituted by two b-jets plus some amount of missing transverse energy.
Optimization cuts were obtained in the context of two scenarios defined by two different values of $\Delta M = M_{\tilde{b}} - M_{\NI}$ to further enhance sensitivity. No excess with respect to the SM prediction is observed and limits on sbottom mass and production cross section are set: $m_{\tilde{b}}$ is excluded up to 230 GeV$/c^2$ at 95$\%$ C.L. for $m_{\NI}$ below 80 GeV$/c^2$ improving the previous CDF results by about 40 GeV$/c^2$.

\section{Non-SUSY searches}
Several SM extensions predict new resonances: at hadron colliders where QCD is overwhelming, lepton-antilepton pair signature has historically been a leading channel for the discovery of new particles because it represents one of the cleanest and most controlled signatures for resonances. An extension of this kind of search is that for other resonances like di-boson resonances which has recently become a reachable goal given the improved analysis capabilities at the Tevatron.

\subsection{Di-lepton high mass search}
The underlying strategy of these searches is to select events with two leptons, construct their invariant mass (that we know accurately) and look for deviations from the expected behaviour in the form of fluctuations especially in the tails of distributions. CDF performed a search for a high-mass neutral resonance decaying in two electrons with 2.5 $\rm{fb}^{-1}$ of data \cite{ref:ee} and observed the largest excess over the SM prediction at $e^{+}e^{-}$ invariant mass of 240 GeV$/c^2$. The probability of observing such an excess arising from fluctuations in the SM anywhere in the range of 150-1000 GeV$/c^2$ is 0.6$\%$. Several models have been tested and SM coupling $Z'$ and the Randall-Sundrum graviton for $k/\bar{M_{Pl}}=0.1$ with masses below 963 and 848 GeV$/c^2$ are respectively excluded at 95$\%$ C.L. 
A similar analysis looking for events with two reconstructed muons in the final state \cite{ref:mm}, has been performed with 2.3 $\rm{fb}^{-1}$ of data. 
In the range of mass considered, the observed width of a dimuon resonance is dominated by the track curvature resolution and results in an approximately constant resolution $\delta m^{-1}_{\mu\mu} $ of 0.17 $\rm{TeV}^{-1}$. For this reason, templates of the $m^{-1}_{\mu\mu}$ variable are constructed, adding the background distributions to the template and comparing the obtained behaviour to the corresponding data distribution for $m_{\mu\mu}> 100$ ${\rm{GeV}}/c^2$.
No significant excess with respect to SM expectation is observed. This allows to set a 95$\%$ C.L. upper limits on $\sigma \times BR(p\bar{p}\rightarrow X \rightarrow \mu\bar{\mu})$ where $X$ is a boson with spin 0, 1 or 2. By using these limits, lower mass limits on sneutrinos in $R_{p}$ violating SUSY models, $Z'$ bosons, and Kaluza-Klein gravitons in the Randall-Sundrum model (see Fig. \ref{fig:gravitonMUMU}), are determined.

\subsection{Di-boson resonance search}
\begin{figure}[ht!]
\begin{minipage}[b]{0.56\linewidth}
\centering
	\includegraphics[width=\linewidth]{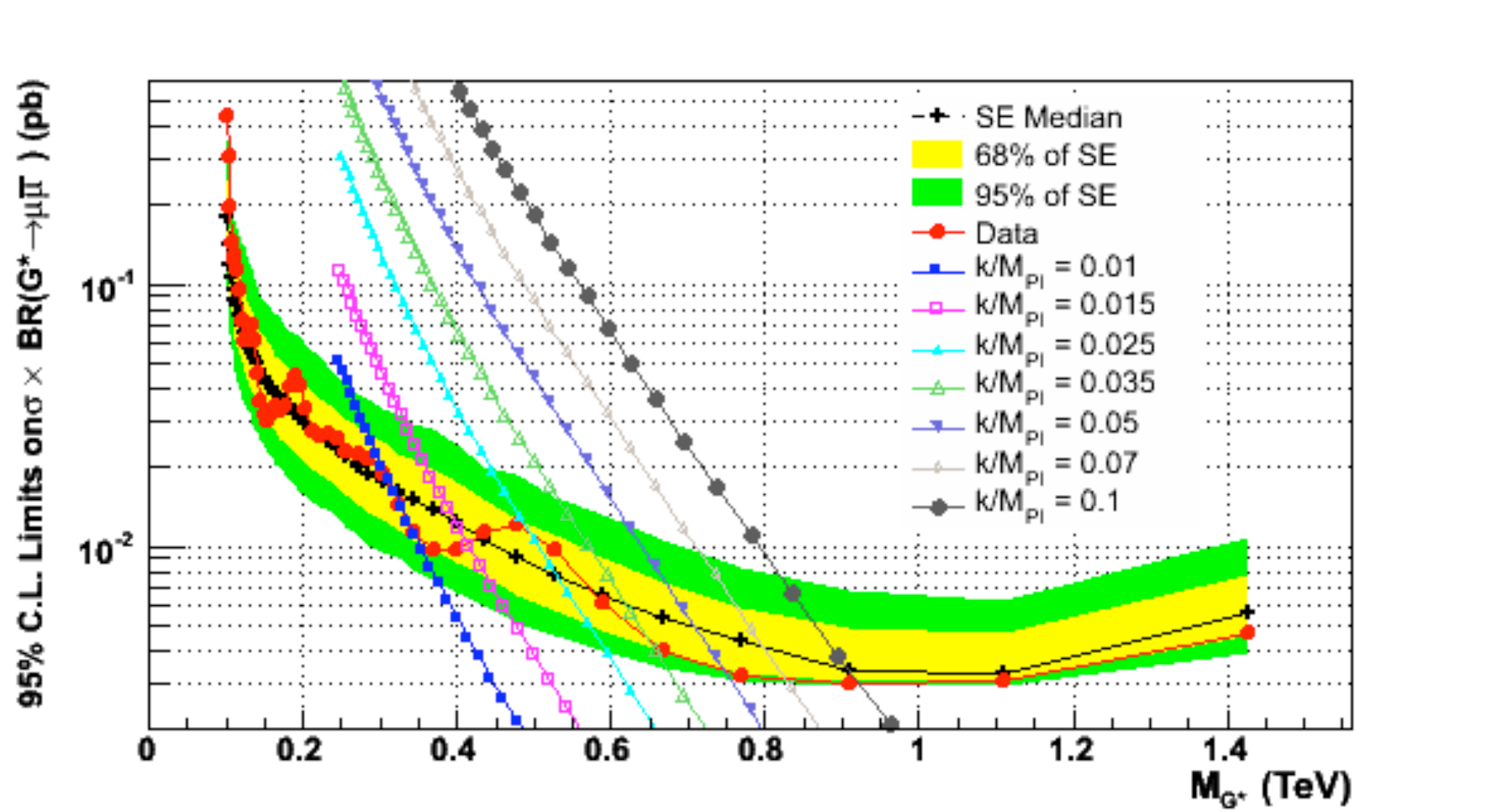}
	\caption{Graviton mass exclusion plot in the search for resonances in di-muon events.}
	\label{fig:gravitonMUMU}
\end{minipage}
\hspace{0.01\linewidth}
\begin{minipage}[b]{0.43\linewidth}
\centering
   \includegraphics[width=\linewidth]{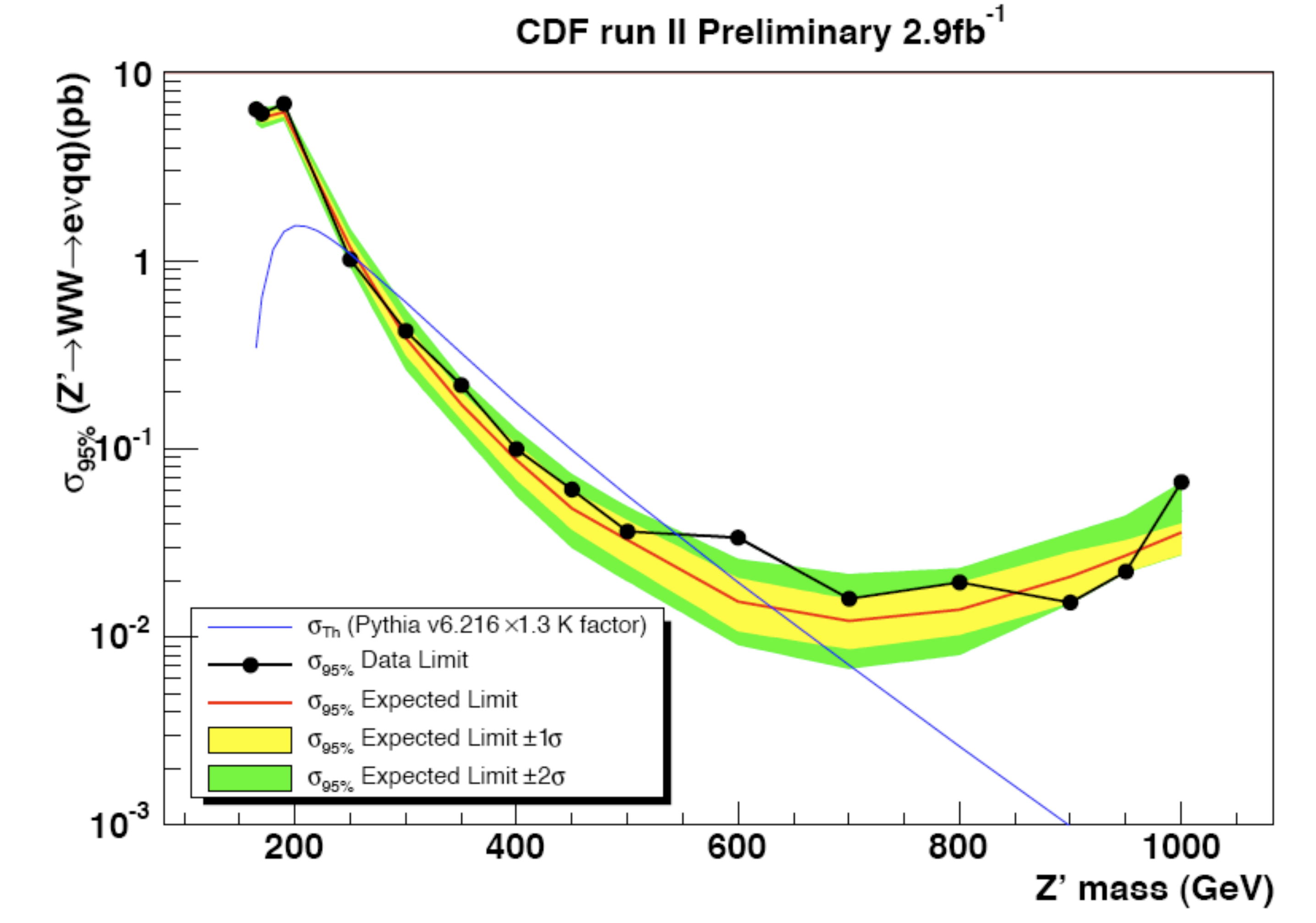}
   \caption{Excluded mass region for a $Z'$ resonance in the di-boson search.}
   \label{fig:diboson}
\end{minipage}
\end{figure}

With 2.9 $\rm{fb}^{-1}$ of data, a search for resonances decaying into a pair of gauge bosons ($W^+W^-$, $W^{\pm}Z^{0}$) was performed at CDF \cite{ref:WZ}.
In this analysis one of the W boson is assumed to decay leptonicaly  in $e \nu_e$ ($BR\simeq10\%$) while the second boson decays into two jets ($BR\simeq70\%$): though background from jets increases, this final state exploits the high hadronic branching ratio of gauge bosons. In the final selection, no excess of data is observed above the expected background, thus results are used to test three resonances hypotheses (R/S graviton, $Z'$, $W'$) and to calculate the corresponding cross section limits at 95 $\%$ C.L. Comparing the limits to their theoretical cross sections, the three mass exclusion regions are also set (see Fig. \ref{fig:diboson} for the specific case of $Z'$).

\section{Conclusions}
CDF has a rich program of searches for new physics. Some recent results are presented in this paper covering supersymmetry and non supersymmetry searches. No excess are found in about 3 $\rm{fb}^{-1}$ of data and several models have been tested and limits set. Analysis tools are in place for updating results with at least twice as much data.

\end{document}